\documentstyle[epsfig,epsf,onecolumn]{mn}

\title[The sdBV star Feige~48]
{Observations of the
pulsating subdwarf B star Feige~48: Constraints on evolution and companions.}

\author[M.D. Reed et al.]
{M. D. Reed,$^1$
S. D. Kawaler$^2$, 
S. Zola$^{3}$, 
X. J. Jiang$^4$, 
S. Dreizler$^5$, 
S.L. Schuh$^5$, \cr 
J.L. Deetjen$^5$,
R. Kalytis$^{6}$,
E. Mei{\v s}tas$^{6}$, 
R. Janulis$^{7}$,
D. Ali{\v s}auskas$^{6}$, 
J. Krzesi{\' n}ski$^{8,9}$,\cr
M. Vuckovic$^2$
P. Moskalik$^{10}$,
W. Og{\l}oza$^{8}$, A. Baran$^8$,
G. Stachowski$^{3,8}$,
D.W. Kurtz$^{11}$,\cr
J.M. Gonz{\'a}lez P{\'e}rez$^{12}$,
A. Mukadam$^{13}$, 
T.K. Watson$^{14}$,
C. Koen$^{13,15}$, 
P. A. Bradley$^{16}$,\cr
M. S. Cunha$^{17}$,
M. Kilic$^{13}$,
E. W. Klumpe$^{18}$, 
R. F. Carlton$^{18}$,
G. Handler$^{15,19}$, \cr
D. Kilkenny$^{15}$, 
R. Riddle$^{2}$, 
N. Dolez$^{20}$, G. Vauclair$^{20}$, M. Chevreton$^{21}$
M. A. Wood$^{22}$, \cr
A. Grauer$^{23}$, 
G. Bromage$^{11}$, 
J. E. Solheim$^{12}$,
R. \O stensen$^{24}$,
A. Ulla$^{25}$,
M. Burleigh$^{26}$, \cr S. Good$^{26}$,
 \"{O}. H\"{u}rkal$^{27}$,
R. Anderson$^{28}$,
and E. Pakstiene$^8$,\\
$^1$ Department of Physics, Astronomy and Material Science, Southwest
Missouri State \\ University, 901 S. National, Springfield, MO 65804 USA 
and visiting astronomer \\ McDonald and Fick Observatories\\
$^2$ Department of Physics and Astronomy, Iowa State University, 
Ames, IA  50011  USA \\
$^{3}$ Astronomical Observatory, Jagiellonian University, ul. Orla 171, 30-244,
Cracow, Poland \\
$^4$ National Astronomical Observatories, Chinese Academy of Sciences, Beijing,
100012, PR China\\
$^5$ Institut f\"{u}r Astronomie und Astrophysik, Universit\"{a}t T\"{u}bingen,
Sand 1, D-72076, T\"{u}bingen, Germany, and  \\
Universit{\"a}ts-Sternwarte G{\"o}ttingen,
 Georg-August-Universit{\"a}t G{\"o}ttingen,
 Geismarlandstra{\ ss}e 11,
 D--37083 G{\"o}ttingen, Germany\\
$^{6}$ Institute of Material Science and Applied Research of Vilnius 
University, Astronomical Observatory, Ciurlionio 29, Vilnius  LT-2009, \\
Lithuania\\
$^{7}$ Institute of Theoretical Physics and Astronomy, Astronomical Observatory,
Go\u{s}tauto 12, Vilnius LT-2600, Lithuania\\
$^{8}$ Mt. Suhora Observatory, Crakow Pedagogical University, ul.
Podchor\c{a}zych 2, PL-30-084 Cracow, Poland \\
$^9$ Apache Point Observatory, P.O. Box 59, Sunspot, NM 88349, U.S.A.\\
$^{10}$ Nicolas Copernicus Astronomical Center, Polish Academy of Sciences, ul.
Bartycka 18, 00-716 Warsaw, Poland \\
$^{11}$ Centre for Astrophysics, University of Central Lancashire, Preston PRI
2HE\\
$^{12}$ Institutt for Fysikk, Universitet i Troms{\o} , N-9037 Troms\o ,
Norway\\
$^{13}$ Department of Astronomy, University of Texas, Austin, TX 78712, USA\\
$^{14}$ Southwestern University, 1001 E. University Avenue, Georgetown, TX
78626, USA\\
$^{15}$ South African Astronomical Observatory, PO Box 9, Observatory 7935,
Cape, South Africa\\
$^{16}$ Los Alamos National Laboratory, X-2, MS T-085, Los Alamos, NM 87545,
USA\\
$^{17}$ Centro de Astrof{\'i}sica da Universidade do Porto, Rua das Estrelas,
4150-762
Porto, Portugal; Instituto Superior da Maia, Av. Carlos \\ de Oliveira Campos, 
4475-690, Avisoso S. Pedro, Castelo da Maia, Portugal\\
$^{18}$ Middle Tennessee State University, Department of Physics and 
Astronomy, Murfreesboro, TN 37132, U.S.A.\\
$^{19}$ Present address: Institut f\"{u}r Astronomie, Universit\"{a}t Wien,
T\"{u}rkenschanzstra\ss e 17, A-1180 Wien, Austria \\
\pagebreak
$^{20}$ Universit\'{e} Paul Sabatier, Observatoire Midi-Pyr\'{e}n\'{e}es, 14
Avenue E. Belin, 31400 Toulouse, France\\
$^{21}$ Observatoire de Paris-Meudon, DAEC, 92195, Meudon, France\\
$^{22}$ Department of Physics and Space Sciences \& SARA Observatory,
Florida Institute of Technology, 
Melbourne, FL 32901-6975, U.S.A.\\
$^{23}$ Department of Physics and Astronomy, University of 
Arkansas at Little Rock, Little Rock, AR 72204, U.S.A.\\
$^{24}$ Isaac Newton Group of Telescopes, E-37800 Santa Cruz de La Palma, Canary
Islands, Spain \\
$^{25}$ Universidade de Vigo, Depto. de Fisica Aplicada, Facultade de Ciencias,
Campus Marcosende-Lagoas, 36200 Vigo, Spain\\
$^{26}$ Department of Physics and Astronomy, University of Leicester, Leicester
LE1 7RH, England\\
$^{27}$ Ege University, Science Faculty, Dept. of Astronomy and Space Sciences,
Bornova 35100 Izmir, Turkey\\
$^{28}$ Department of Physics and Astronomy, University of North Carolina,
Chapel Hill, NC 27599-3255, USA 
}

\date{Accepted     
      Received }

\begin{document}

\maketitle

\begin{abstract}
Since pulsating subdwarf B (sdBV or EC14026) stars were first discovered
(Kilkenny et al, 1997), 
observational efforts have tried to realize their potential for
constraining the interior
physics of extreme horizontal branch (EHB) stars. 
Difficulties encountered along the way include uncertain mode 
identifications and a lack of stable pulsation mode properties.
Here we report
on Feige~48, an sdBV star for which follow-up observations have
been obtained spanning more than four years, which shows some
stable pulsation modes.

We resolve the temporal spectrum into
five stable pulsation periods in the range 340 to 380 seconds
with amplitudes less than 1\%, and
two additional periods that appear in one dataset each. The three
largest amplitude periodicities are nearly equally spaced, and
we explore the consequences of identifying them as a rotationally
split $\ell=1$ triplet by consulting with a representative stellar
model. 
 
The general stability of the pulsation amplitudes and phases allows us to
use the pulsation phases to constrain the timescale
of evolution for this sdBV star.   Additionally, we are able 
to place interesting limits on any stellar or planetary
companion to Feige~48.

\end{abstract}

\begin{keywords}

Stars: oscillations -- stars: variables -- 
stars: individual (Feige~48)

\end{keywords}

\section{Introduction}
To date, over 30 pulsating subdwarf B (EC~14026 or sdBV) stars
 have been identified, with
pulsation periods ranging from 68 to 528 seconds and with amplitudes generally
less than 50 millimagnitudes (mmag). 
For recent reviews of this class of stars, see
Kilkenny (2001), and Reed, Kawaler \& Kleinman (2000),
for observational properties;
Charpinet, Fontaine \& Brassard (2001 and references therein) describe
in detail some important aspects of pulsation theory in sdB stars.
Most sdBV stars show periods at the short end of the range, and probably 
represent stars close to the Zero Age Horizontal Branch (ZAHB).  PG~1605+072
is the longest period, and lowest gravity, sdBV star, with Feige~48 being 
an intermediate object.  In general, the longer-period sdBV stars 
represent more highly evolved objects.

Feige~48 was identified as a ``faint blue star'' as part of the 
Feige survey (Feige, 1958). It was re-categorized as an sdB star
when it was observed as part of the Palomar-Green survey (Green,
Schmidt, \& Liebert, 1984).   Koen et al. (1998; hereafter K98) 
identified five pulsation periods in Feige~48 in 
six observing
runs from 1997 May to 1998 February. The periods detected by K98
range from 342 to
379 seconds with the largest amplitude being 6.4 mmag.  Amplitude variability
led K98 to conclude that mode beating was probably
present implying that other unresolved modes were present in their
data. This provided the motivation for our follow-up observations.
Heber, Reid, \& Werner (2000; hereafter HRW) 
obtained a high resolution (0.09\AA) 
spectrum of Feige~48, from which they determined
log $g$=5.50$\pm$0.05 and T$_{\rm eff}$=29500$\pm$300 K. This places
Feige~48 among the coolest sdBV stars known with a surface gravity
intermediate between PG~1605+072 and the
rest of the class.

Here we report on our multi-year campaign of high-speed photometry of Feige~48.
  In Section~2, we outline our observations.  Section~3 describes the 
time series analysis and period identifications.  We report on a 
stellar model fit to 
Feige~48 in Section~4.  The phase stability of pulsations is 
described in Section~5, where we use this stability to place interesting 
limits on any possible planetary companion. Section 6 gives our conclusions 
and outlines future observations for Feige~48.

\section{High Speed Photometry}
\begin{table*}
\centering
\caption{Observations of Feige~48 \label{tab01}}
\begin{tabular}{|lcrl|lcrl|} \hline
Run & Length &
Date & Observatory & Run & Length & Date & Observatory\\
 & (hrs) & UT & & & (hrs) & UT &  \\ \hline
tex-007 & 1.2 & 1997.03.05 & McDonald 0.9m & suh-106 & 3.2 & 2000.08.11 & Suhora
0.6m\\
tex-018 & 2.1 & 1997.02.06 & McDonald 0.9m & jxj-125 & 2.5 & 2000.26.11 & BAO
0.85m\\
tex-223 & 5.7 & 1998.23.01 & McDonald 0.9m & jxj-128 & 3.2 & 2000.27.11 & BAO
0.85m\\
tex-236 & 2.9 & 1998.28.01 & McDonald 0.9m & suh-107 & 8.8 & 2000.21.12 & Suhora
0.6m\\
tex-239 & 5.5 & 1998.29.01 & McDonald 0.9m & suh-108 & 3.4 & 2000.22.12 & Suhora
0.6m\\
tex-241 & 5.9 & 1998.30.01 & McDonald 0.9m & mdr145 & 8.0 & 2001.18.01 & Fick
0.6m \\
tex-246 & 6.7 & 1998.01.02 & McDonald 0.9m & mdr146 & 8.2 & 2001.20.01 & Fick
0.6m \\
 mdr006 & 2.2 & 1998.22.11 & McDonald 0.9m &mdr147 & 3.6 & 2001.21.01 & Fick
0.6m\\
 mdr009 & 2.6 & 1998.23.11 & McDonald 0.9m &mdr148 & 9.2 & 2001.22.01 & Fick
0.6m\\
 mdr012 & 2.6 & 1998.24.11 & McDonald 0.9m &mdr149 & 9.5 & 2001.24.01 & Fick
0.6m\\
 mdr017 & 2.8 & 1998.26.11 & McDonald 0.9m &mdr150 & 9.2 & 2001.25.01 & Fick
0.6m\\
 mdr018 & 6.0 & 1999.06.03 &   McDonald 2.1m &mdr151 & 7.3 & 2001.01.02 & Fick
0.6m\\
 mdr021 & 5.5 & 1999.09.03 &   McDonald 2.1m &asm-0086 & 4.4 & 2001.19.04 &
McDonald 0.9m\\
 mdr023 & 4.1 & 1999.10.03 &   McDonald 2.1m &sara0082 & 3.7 & 2001.21.04 & SARA
0.9m\\
 mdr24a & 6.0 & 1999.11.03 &   McDonald 2.1m &tkw-0065 & 7.3 & 2001.22.04 &
McDonald 0.9m\\
 mdr29a & 8.7 & 1999.15.03 &   McDonald 2.1m &sara0086 & 6.8 & 2001.24.04 & SARA
0.9m\\
 mdr030 & 1.5 & 1999.17.03 &   McDonald 0.9m &IAC80A08 & 0.8 & 2001.25.04 &
Teide 0.8m\\
 mdr033 & 2.0 & 1999.19.03 &   McDonald 0.9m &sara0088 & 7.0 & 2001.25.04 & SARA
0.9m\\
 mdr035 & 6.0 & 1999.20.03 &   McDonald 0.9m &IAC80A09 & 6.1 & 2001.26.04 &
Teide 0.8m\\
 mdr039 & 5.3 & 1999.23.03 &   McDonald 0.9m &sara0089 & 5.4 & 2001.26.04 & SARA
0.9m\\
 caf48r1r2 & 7.5 & 1999.12.04 & Calar Alto 1.2m &suh-102 & 3.9 & 2001.29.04 &
Suhora 0.6m\\
 suh-75 & 1.7 & 1999.13.04 &   Suhora 0.6m  &suh-103 & 0.6 & 2001.30.04 & Suhora
0.6m\\
 caf48r3 & 9.3 & 1999.13.04 & Calar Alto 1.2m &suh-104 & 0.1 & 2001.30.04 &
Suhora 0.6m\\
 mdr091 & 4.0 & 1999.10.12 & Fick 0.6m &IAC80A17 & 6.1 & 2001.30.04 & Teide
0.8m\\
 mdr093 & 6.2 & 1999.13.12 & Fick 0.6m &mdr198 & 7.0 & 2002.17.02 & Fick 0.6m \\
 mdr095 & 2.4 & 1999.14.12 & Fick 0.6m &mdr199 & 1.5 & 2002.05.04 & Fick 0.6m \\
 mdr096 & 4.9 & 1999.16.12 & Fick 0.6m &mdr200 & 4.5 & 2002.06.04 & Fick 0.6m\\
 mdr098 & 2.1 & 2000.08.02   & McDonald 0.9m &suh-109 & 1.9 & 2002.07.05 &
Suhora 0.6m\\
 mdr100 & 1.3 & 2000.08.02   & McDonald 0.9m &sara141  & 7.3 & 2002.07.05 & SARA
0.9m\\
 mdr103 & 4.6 & 2000.10.02   & McDonald 0.9m &suh-110 & 5.4 & 2002.08.05 &
Suhora 0.6m\\
 mdr108 & 2.7 & 2000.12.02    & McDonald 2.1m &suh-111 & 1.2 & 2002.09.05 &
Suhora 0.6m\\
 mdr110 & 1.0 & 2000.12.02   & McDonald 2.1m &adg-519 & 0.7 & 2002.11.05 &
Mt.Bigelow 1.5m\\
 mdr111 & 3.3 & 2000.28.02   & Fick 0.6m &suh-112 & 1.2 & 2002.12.05 & Suhora
0.6m\\
 mdr112 & 3.7 & 2000.01.03   & Fick 0.6m &fe0512oh & 2.2 & 2002.12.05 & OHP
1.9m\\
 mdr113 & 4.0 & 2000.02.03    & Fick 0.6m & jr0512 & 3.4 & 2002.12.05 & Moletai
1.65m\\
 mdr114 & 1.5 & 2000.03.03   & Fick 0.6m & suh-113 & 4.4 & 2002.13.05 & Suhora
0.6m\\
 mdr115 & 6.3 & 2000.04.03   & Fick 0.6m &  jr0513 & 3.7 & 2002.13.05 & Moletai
1.65m\\
mdr116 & 2.6 & 2000.04.03 & Fick 0.6m & fe0513oh & 1.1 & 2002.13.05 & OHP 1.9m\\
mdr117 & 6.0 & 2000.05.03 & Fick 0.6m & jgp0209  & 0.7 & 2002.14.05 & Teide
0.8m\\
mdr118 & 3.3 & 2000.05.03 & Fick 0.6m & jkt-003 & 1.7 & 2002.14.05 & JKT 1.0m\\
mdr119 & 3.8 & 2000.04.05 & Fick 0.6m & jkt-007 & 5.6 & 2002.17.05 & JKT 1.0m\\
mdr120 & 6.0 & 2000.05.05 & Fick 0.6m &  a0239 & 1.0 & 2002.18.05 & McDonald
2.1m\\
suh-101 & 6.0 & 2000.02.11 & Suhora 0.6m& jr0518 & 1.7 & 2002.18.05 & Moletai
1.65m\\
suh-102 & 2.9 & 2000.03.11 & Suhora 0.6m& jr0519 & 2.4 & 2002.19.05 & Moletai
1.65m\\
suh-103 & 1.4 & 2000.05.11 & Suhora 0.6m& jr0520 & 2.2 & 2002.20.05 & Moletai
1.65m\\
suh-104 & 9.7 & 2000.05.11 & Suhora 0.6m& jr0521 & 3.0 & 2002.21.05 & Moletai
1.65m\\
suh-105 & 4.6 & 2000.07.11 & Suhora 0.6m &       &     &         &      \\
\hline
\end{tabular}
\end{table*}

We began observing Feige~48 in 1998 November, with our
most recent observations acquired in 2002 May.  Table 1 provides a complete
list of our observations (including K98's observations).
We acquired most of the data using 3 channel 
photoelectric photometers as described in Kleinman, Nather, \& 
Phillips (1996).
At Fick Observatory, we used a 2 channel
photometer of similar design. The Fick
data typically have an $\approx$30 minute gap during the night, as the
telescope mount requires the photometer to be disconnected when the telescope
exchanges sides on the pier. Because this instrument is a 2 channel
photometer, the data were occasionally interrupted to measure the sky
background. All
photometers used Hamamatsu R647 photomultiplier tubes. Data acquired at
Calar Alto and SARA
were obtained using CCDs with 5 second exposures on a 30 and 15 second 
duty  cycle, respectively.  As both the target 
and comparison star were in the same CCD field, differential 
photometry removed extinction and sky variation. However, since 
extinction is wavelength--dependent, colour differences between the stars 
produced small non-linear trends in the data.  We remove these trends
by dividing by a low (2 - 4) order polynomial fit to the 
single--night data. Bad points in the CCD
data were removed by hand.
 No filters were
 used during any of the photoelectric observations
to maximize the photon count rate, whereas the CCD observations used filters
to approximate the passband of the Hamamatsu phototubes (Kanaan et al.
2000).

As Table 1 indicates, we have obtained a total of $\approx$380 hours of time
series photometry 
on Feige 48.   The data span from 1998 January to 2002 May (the
two  runs in 1997 were too short and too temporally separated
to be useful for this analysis).

\section{The Pulsation Spectrum of Feige~48}

We follow the standard procedure for determining pulsation frequencies
from time series photometry obtained using the Whole Earth Telescope  (see, for
example, O'Brien et al. 1998).  
In short, we identify the principal pulsation 
frequencies with a Fourier transform of light curves of individual nights.  We 
then combine data from several contiguous nights to refine the frequencies.  
Once the main frequencies are found, we then do a nonlinear least squares 
fit for 
the frequencies, amplitudes, and phases of all identified peaks, along with 
their uncertainties.

\begin{table}
\centering
\caption{Subgroups used in pulsation analysis. \label{tab02}}
\begin{tabular}{ccc}\hline
Group & Inclusive dates & Hours of data. \\ \hline
I & 1998.28.01 - 1998.01.02 & 26.7\\
II & 1998.22.11 - 1998.26.11 & 10.2\\
III & 1999.06.03 - 1999.13.04 & 63.6\\
IV & 1999.10.12 - 1999.16.12 & 17.5\\
V & 2000.08.02 - 2000.05.03 & 42.4\\
VI & 2000.04.05 - 2000.05.05 & 9.8\\
VII & 2000.02.11 - 2000.22.12 & 45.7\\
VIII & 2001.18.01 - 2001.01.02 & 55.0\\
IX & 2001.19.04 - 2001.30.04 & 52.0\\
X & 2002.05.04 - 2002.21.05 & 56.8\\ \hline
\end{tabular}
\end{table}

\begin{figure*} \centering
\centerline{\epsfig{figure=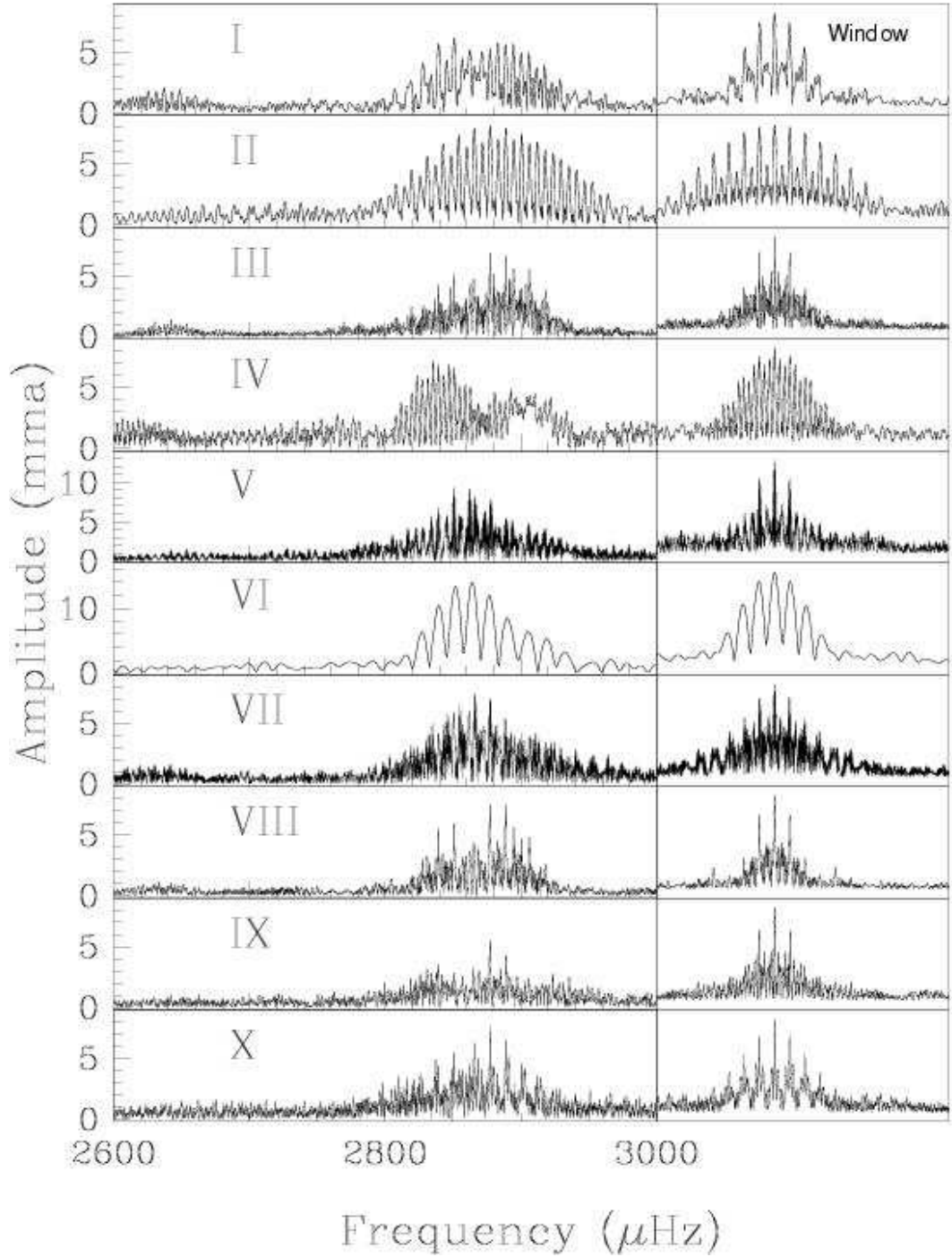,width=5.5in}}
\caption{Temporal spectra and Window functions of Feige~48 for
the groups listed in Table~\ref{tab02}. }
\label{grp1a}
\end{figure*}

\begin{figure*}
\epsfig{figure=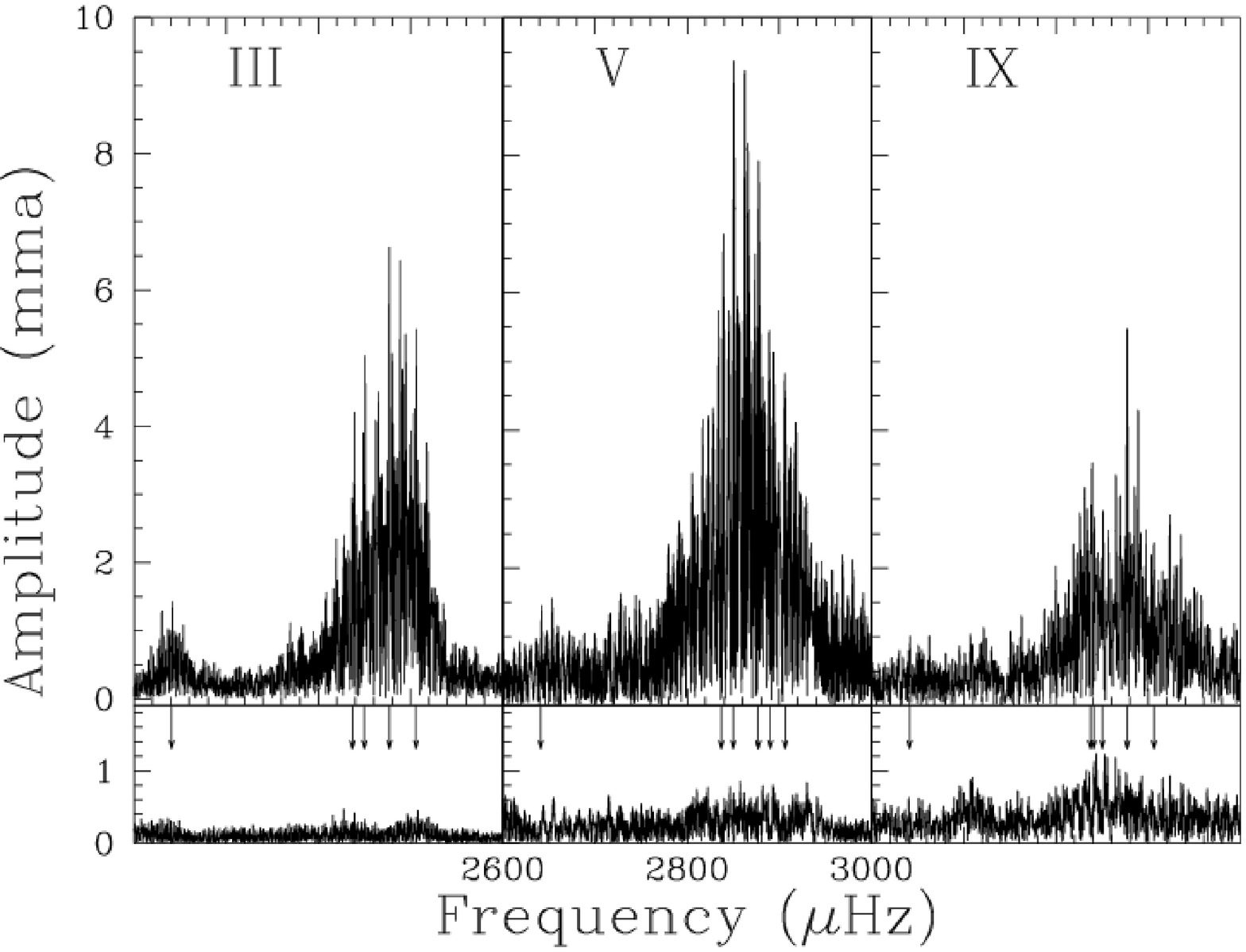,width=\textwidth}
\caption
{Temporal spectra (top) and residuals after prewhitening by frequencies
in Table~\ref{tab03} (bottom).  \label{fig01} }
\end{figure*}

To work around monthly and annual
 gaps between observing runs, we begin our analysis of the
data in separate, relatively contiguous subgroups.
The dates and data hours obtained for the subgroups are given in
Table~\ref{tab02}, the temporal spectra and window functions
 for the groups are plotted in Fig.~\ref{grp1a}. All groups were
analysed independently, without using periods detected in other groups. 
This independent group reductions decrease the likelihood of selecting
a daily alias over a real pulsation.  
 Only in our Group X data, do we
detect a mode ($f$2) inconsistent with the other group reductions. As such,
we presume that our periods for $f1-f5$ are not aliases, with the exception
of $f$2 in Group X, which is a daily alias away from the real period.
Frequencies determined for the better
data sets  are in Table~\ref{tab03} with the 
corresponding temporal spectra of the best groups (Groups III, V, and IX) and 
the prewhitened residuals in Fig.~\ref{fig01}.
Though some signal remains in the Fourier transform 
after prewhitening within these groups, 
we cannot distinguish any remaining pulsation frequencies from noise.

From our
best solution for the Group III data, temporally
adjacent groups were added one at a time and frequencies and amplitudes
were fit again until a satisfactory fit was determined for all the data.
Though there is still a chance that some modes may be off by an annual or
monthly alias, the
periods, frequencies and amplitudes in Table~\ref{tab04}
 represent our best solution. Also indicated by Table~\ref{tab03} are three
pulsation periods that appear only within the span of a
single group. However, the amplitudes are sufficiently low that
some possibility exists that
they could be due to noise and\/ or aliasing.  As such, we will not include them
in our analysis, but we mention them as they are possibly stochastically
excited modes within an otherwise stable pulsation spectrum.
Note also that an error corresponding to an annual (or even  monthly) alias
produces only a small change in the period or frequency, and so will not
affect the modeling for asteroseismic analysis.
Such a mistake, however, would be fatal to the
period stability analysis.

\begin{table*}
\centering
\caption{Comparison of the pulsation frequencies (in $\mu$Hz) detected in
various runs. 
Formal least squares errors are provided in parentheses. \label{tab03}}
\begin{tabular}{lllllllll} \hline
Group & $f1$ & $f2$ & & $f3$ & & 
$f4$ & &$f5$ \\ \hline
$\mu$Hz \\ \hline
I$^{\dag}$ & 2636.96(15) & & & 2850.530(40) & 2874.40(23) & 2877.310(50)& &
2917.700(70)$^{\star}$ \\
III & 2641.98(1) & 2837.53(1) & & 2850.833(4)& & 2877.157(3) & &2906.275(4) \\
V & 2641.49(2) & 2837.53(1)  & & 2850.833(3) & & 2877.177(4) &2890.025(19)
&2906.299(8)  \\
VIII & 2642.00(7) & 2837.53(5) & & 2850.818(17) & & 2877.153(12)&           &
2906.266(22) \\
IX & 2641.98(6) & 2837.78(3) & 2841.151(9) & 2850.946(26) & & 2877.185(13)& &
2906.144(23) \\ 
X & 2641.86(6)  & 2826.97(6)$^{\star}$ & & 2850.811(12) & & 2877.220(10) &
&2906.665(43)\\ \hline
\multicolumn{8}{l}{$^{\dag}$ These frequencies are directly from K98.}\\
\multicolumn{8}{l}{$^{\star}$ Indicates modes offset by approximately the daily
alias (11.56 $\mu$Hz).} \\ \hline
\end{tabular}
\end{table*}

\begin{table}
\centering
\caption{Least squares solution to the entire data set. Formal least squares
errors are
in parentheses.  \label{tab04}}
\begin{tabular}{clll}
Mode & Period & Frequency & Amplitude \\ 
& (sec) & ($\mu$Hz) & (mma) \\ \hline
$f1$ & 378.502960(37) &   2641.98731(23) &   1.19(6) \\
$f2$ & 352.409515(32) &   2837.60791(21) &   1.30(6)\\
$f3$ & 350.758148(3) &   2850.96729(7)  &  4.17(6) \\
$f4$ & 347.565033(19) &   2877.15942(4)  &  6.35(6) \\
$f5$ & 344.082794(15) &   2906.27734(8)  &  3.57(6) \\ \hline
\end{tabular}
\end{table}

Feige~48 shows five distinct and consistent
 pulsation modes.
Four of the stable modes 
 cluster with periods near 350 seconds and a single mode lies at
378.5 seconds. The three shortest period modes also have the highest
amplitudes -- over three times higher than the two longer period
modes. 

\begin{figure*}
\centerline{\epsfig{figure=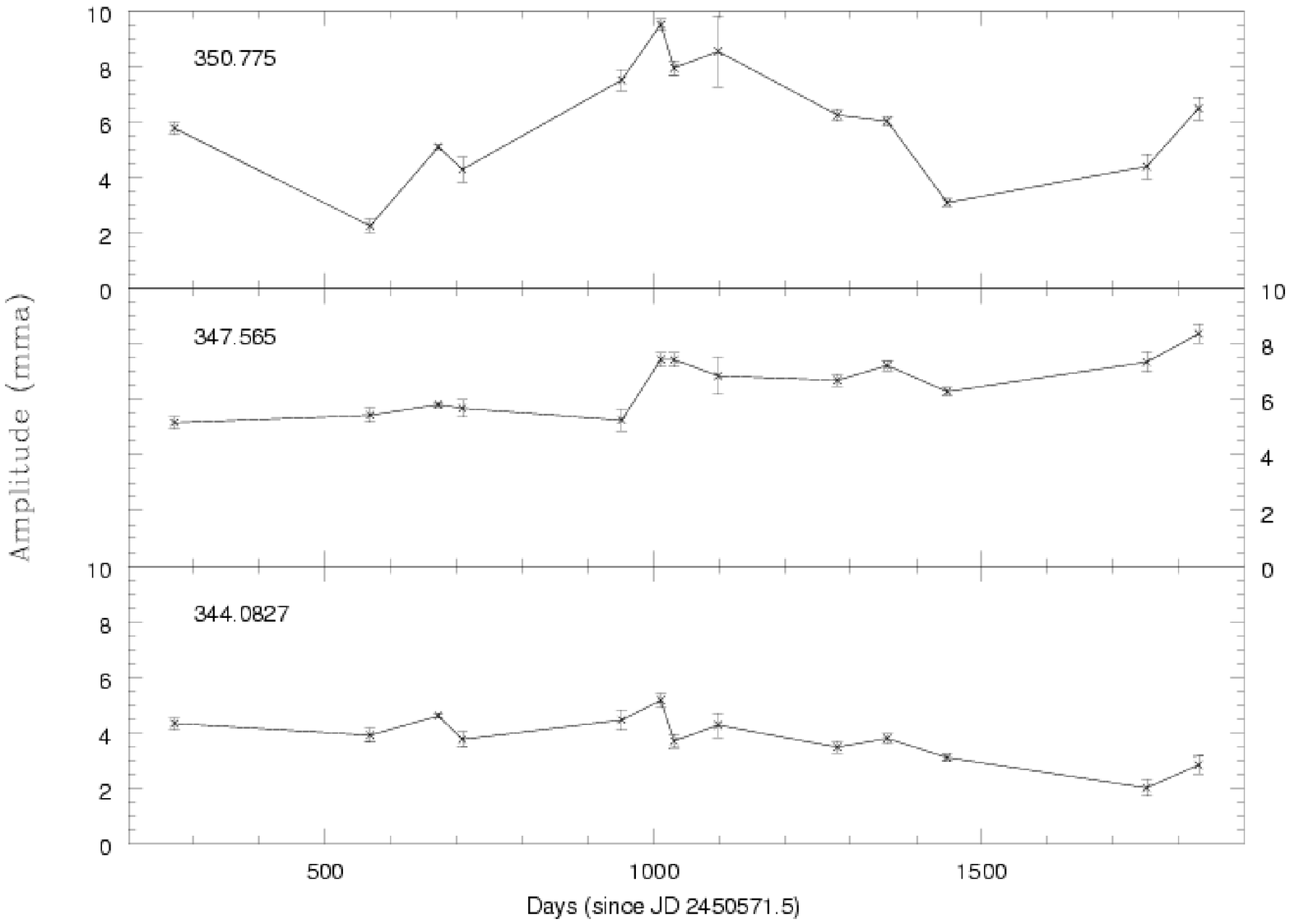,width=5.5in}}
\caption{Pulsation amplitudes for the 3 largest amplitude modes.}
\label{amppic}
\end{figure*}

Our initial interest in follow-up observations of Feige~48 was amplitude
variability observed by K98.  Our hope was to detect unresolved pulsations that
could be responsible for the apparent amplitude variability they reported.
However, it appears that the pulsations are resolved, and each has
a variable amplitude of at least 30 \%. The amplitudes determined for each
of our groups for the three high amplitude modes are shown in 
Figure~\ref{amppic}. All 3 amplitudes show change,  though only
$f3$ has a dramatic degree of variability.

\section{Analysis of the pulsation spectrum of Feige~48}

\subsection{Frequency splittings in Feige~48}

For most sdBV stars, sufficient data do not exist to resolve the complete
pulsation structure.  For stars with resolved temporal spectra
(Kilkenny 2001) there are typically too many modes to be accounted for by
current pulsation theory unless high $\ell$ values are included (where, by high
$\ell$, we mean $\ell \ge 3$).  Though higher $\ell$ modes could be present, such
modes suffer from severe cancellation effects across the unresolved stellar
disk.  In any case, identification of rotationally split multiplets ($\ell=1$
triplets or $\ell=2$ quintuplets of nearly equally spaced modes, for example)
could aid in accounting for the many modes seen.  Unfortunately, previous
studies have been limited by a lack of equally spaced (in frequency
or period) pulsations as a constraint on the observed $\ell$ values (with the
exception of PG~1605+072; Kawaler 1999).

Even though Feige 48 has a relatively simple pulsation spectrum,
understanding why this star pulsates with these frequencies still presents a
problem.  
The tight cluster of four modes with periods spanning a range of less than 10
seconds is impossible to accommodate with purely radial pulsation modes in sdB
models.  Even appealing to nonradial pulsations, the closeness of these periods
means that rotation (or other departures from spherical symmetry) must play a
role.  The reason is that if all are $m$=0 modes, standard
evolutionary models of sdB stars at 
the same $T_{\rm eff}$ and $\log g$ as Feige
48  do not have dense enough frequency spectra to account for these 4 modes,
even if one appeals to modes of higher degree than $\ell=3$.

However, Feige~48 may provide important clues in its observed pulsation
spectrum.  The frequency difference between $f3$ and $f4$ (26.2 $\mu$Hz) is
very close to the difference between $f4$ and $f5$ (29.1 $\mu$Hz).  Though
not exact, this splitting is highly suggestive that $f3$, $f4$, and $f5$
are a rotationally split mode, and probably $\ell=1$.  We note that the
frequency splitting is not precisely equal (26 versus 29 $\mu$Hz).  First
order pulsation theory (if applicable here) says that the splitting should
be precisely equal -- though observations of rotational splitting in white
dwarf stars show asymmetries such as this (i.e. in PG~2131+066, Kawaler et al.
1995 or GD358, Winget et al. 1994). Such asymetries can be caused by
many higher-order processes including rotation and magnetic fields
However this mainly observational
 paper is not the medium for such a discussion, so we will ignore the small
departures from symmetry.

Another possible clue is 
that the spacing between $f2$ and $f3$ is 13.2 $\mu$Hz,
which is almost exactly half the frequency
spacing between $f3$ and $f4$.  Thus we could
interpret the observed frequencies as follows: the $f3$, $f4$, $f5$ set make an
$\ell=1$ triplet, meaning the model needs to fit $f4$ with an $\ell=1$, and $f3$
and $f5$ reveal the rotation rate.  Or, $f2$, $f3$, and $f4$ are three
components of an $\ell=2$ quintuplet with a spacing of $13.2$ $\mu$Hz. In this
case the $m=0$ component could either be $f3$ itself or $f3 + 13.2$ $\mu$Hz
(with a period of 349.1 seconds).

\subsection{Comparison with standard evolutionary models}

Though we by no means have a complete grid of models and cannot quantify
the uniqueness of our results, we can see if either of the above
possibilities is consistent with expectations from standard
evolutionary models of sdB stars. We take the approach of Kawaler (1999);
create several series of evolutionary models that pass through the
spectroscopic errorbox of  $T_{\rm eff}$
and $\log g$, being sure to sample several regions. From these grids,
we search for periods near those observed, creating a list of models and
pulsation periods. From within this list, 
we further constrain the match using the required value of $\ell$
and $m$ for each possible interpretation of the splittings noted above. For
the best model series, we create models with smaller evolutionary steps,
finding the one with the best fit to the periods.

With 5 frequencies available, there are multiple ways to match observations
with model frequencies depending on the assumed values of $\ell$, and $m$
for each mode. The most obvious assumption to make is that modes $f3$,
$f4$, and $f5$ form a rotationally split triplet with $\ell$=1,
$m$=-1,0,+1.  This choice has no ``missing members'' of the multiplet. With
this assumption, the model that fits the spectroscopic data must show an
$\ell=1$ mode at $f4$, and modes at $f1$ and $f2$. Other choices for this
triplet which \emph{do} require that some components of the multiplet are
unobserved are  $\ell$=2, with $\Delta m = 2$ ($m$=-2,0,+2) or $\Delta m =
1$ (e.g. $m$=-2,-1,0 or $m$=-1,0,+1). Similarly, the interpretation of $f2$,
$f3$, and $f4$ being components of an $\ell=2$ multiplet, as described above, is
viable.   

Without choosing a priori one of these interpretations, we examined
evolutionary models within our preliminary grid as described below.  We
generated full evolutionary stellar models using a version of the ISUEVO
stellar evolution program (Dehner 1996; Dehner \& Kawaler 1995; Kawaler
1999) that incorporates semiconvection in the core helium--burning phase.
Our initial grid of evolutionary tracks and models spans the observed range
of T$_{\rm eff}$ and $\log g$ for sdB stars with a core mass of
0.47M$_{\odot}$ and solar metallicity. We computed evolutionary tracks for
models with hydrogen envelope masses ranging from 0 to 0.00550 M$_{\odot}$.
From this initial grid, we focused on model series
 whose evolutionary tracks pass
within 1$\sigma$ of the spectroscopically determined values of T$_{\rm
eff}$ and $\log g$ (HRW).   For these models, we then calculated their
pulsation periods to see which (if any) had radial or nonradial pulsation
periods near those observed, for any possible choice of $\ell$ and
identification of the $m=0$ component of a rotationally split multiplet. 
For the sequence that most closely matched the observed periods, 
we iteratively produced models with smaller differences in H shell masses
and smaller evolutionary time-steps.

\begin{figure*}
\centerline{
\epsfig{figure=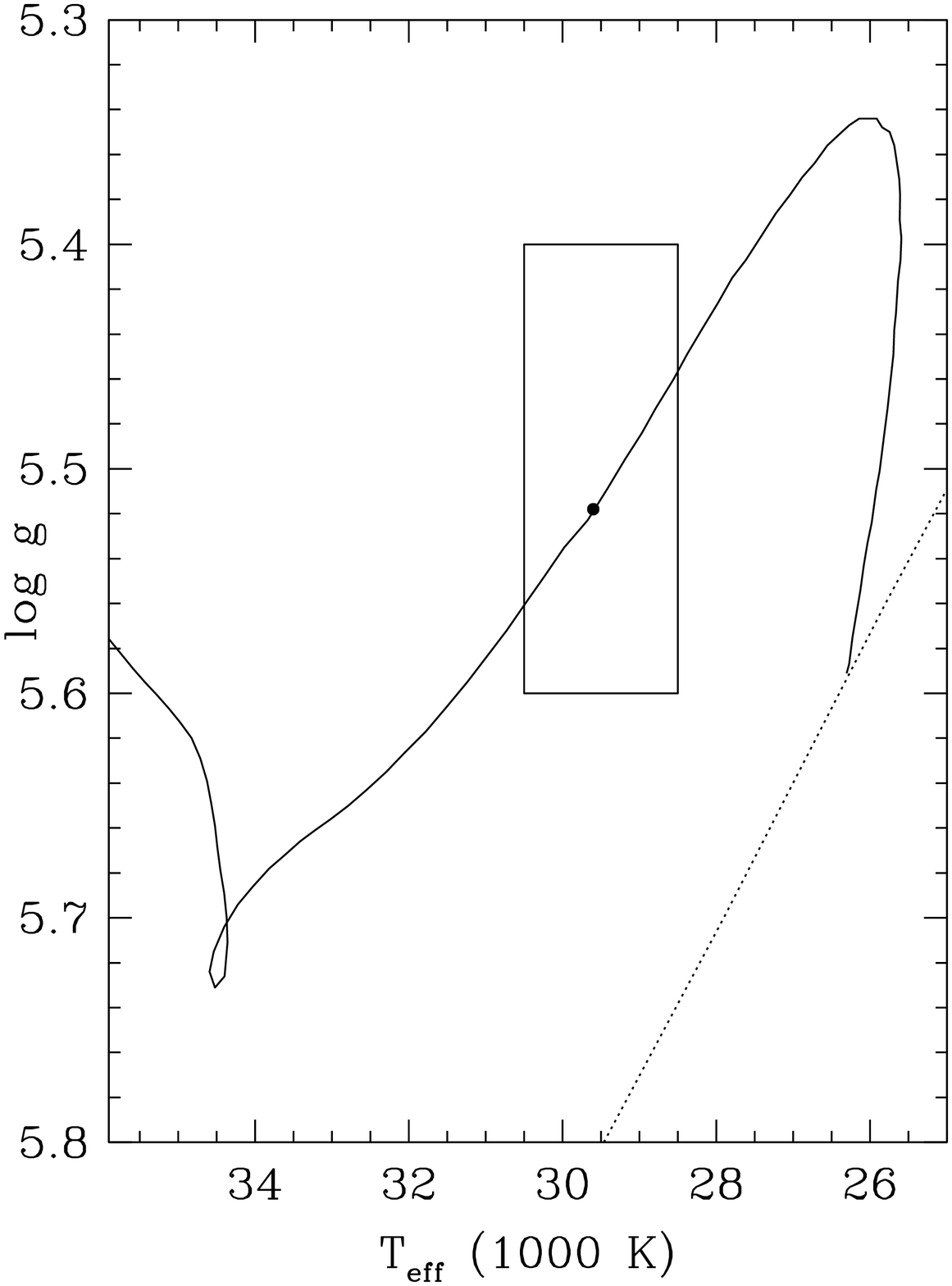,width=3.0in}\epsfig{figure=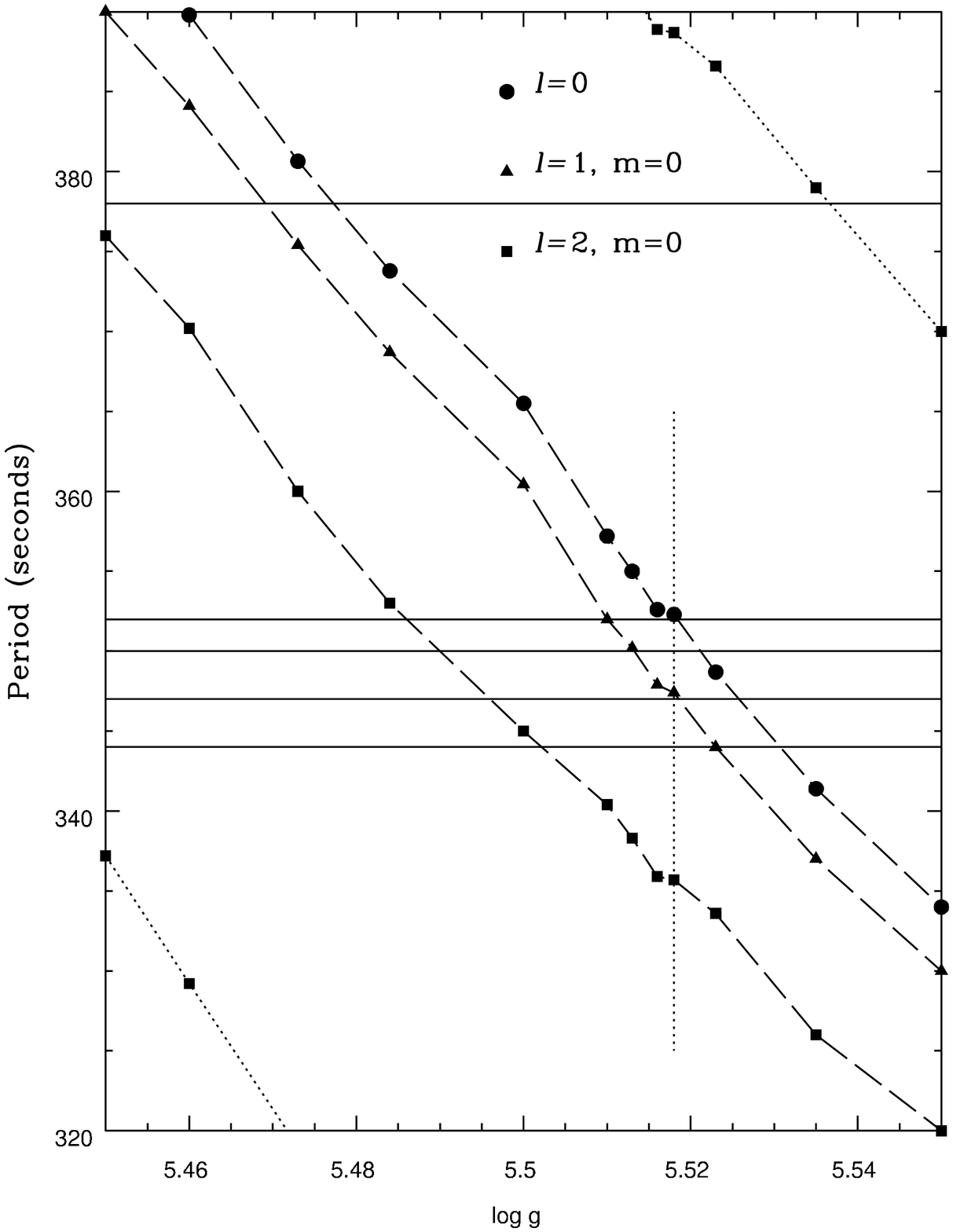,width=3.0in} }
\caption
{Comparison of the model to the observations. The 
left panel shows the evolutionary
model track (solid line) containing our best-fitting model (dot). The dashed
line is the ZAHB and the rectangle is the spectroscopic 1$\sigma$ error box.
The right panel compares the model periods (points) to those observed
(solid lines). The vertical dashed line indicates the best-fitting model.
\label{fig03} }
\end{figure*}

We did find a model that matched the spectroscopic constraints, and,
with an ad hoc rotational splitting of 27.73$\mu$Hz (the average of the 
observed splittings),
could explain four of the five observed frequencies. The closest model
fit came from a model in the evolutionary track shown in
Fig.~\ref{fig03}.  This model does an excellent job in explaining $f2$
through $f5$ and requires the identification of $f3$, $f4$, and $f5$ as
a rotationally split $\ell=1$ triplet.  The model is fairly evolved (as
we suspected) and has nearly exhausted its core helium, with only
0.74\% (by mass) of the core composed of helium.  Table~\ref{table12}
shows a comparison between the observed periods and best-fitting model
periods, and the observed spectroscopic properties and the model
parameters. This
model fits all but the lowest frequency to a precision
of better than 0.2\,s. 
The model temperature agrees to within 150~K, and log $g$
agrees to within 0.02 dex of the measured values; these are well within the
$1\sigma$ uncertainties quoted by HRW.

Despite the perception that there are many degrees of freedom, this
procedure produced only 1 model series that fit the observations: i.e. no
models had appropriate $\ell$=2 or $\ell$=1, $n$=2 periods. Of course,
as we do not have model grids covering all parameter space, it is possible
another model, with perhaps a different core mass or metalicity may fit
better or with altogether different $\ell$ identifications.
This procedure also failed to produce a model that could explain all 5 observed
frequencies in terms of normal modes and rotational splitting. Though the
model does produce an $\ell$=3, $m$=0 mode at 374\,s period, which is near
the observed 378\,s mode, we currently find no evidence to suggest that Feige~48
 has $\ell
>2$. As such, we must confess
that our model does not reproduce this period without appealing       
to high $\ell$. Without further evidence (such as other observed    
members of the multiplet) for invoking high $\ell$ modes, we are forced
to leave $f1$ as unmatched by our model.
Additionally, any $\ell$=2 matches fitted less observed periods. 
Since we do not have a complete
sample of models and this is really just an illustrative example, we are
not alarmed. However, it could also indicate that our current
models do not include enough physics to be accurate.

The pulsation results for the closest-fitting modes in our best-fitting model 
series are shown in Fig.~\ref{fig03}. Even a  small change in log $g$ (as an
indication of age) of 0.005 dex changes the calculated periods by more than 3 seconds, 
worsening the fit to the observations. Likewise, a change in the
envelope layer thickness quickly destroys the fit by moving the evolutionary
track's path away from the spectroscopic error box.  As the right panel of 
Fig.~\ref{fig03} shows, within this period space, model periods are relatively
uncrowded. Overtones are separated by $\sim$100~s for $\ell$=0 and 1 modes, and
$\sim$50 seconds for $\ell$=2 modes. Overtones for $\ell$=2 do appear in the
top right and lower left of the figure.

\begin{table}
\centering
\caption{Comparison of observations with our
best-fitting evolutionary model.}
\label{table12}
\begin{tabular}{ccccccc} \hline
&  Frequency & \multicolumn{2}{c}{Period (sec)} & \multicolumn{3}{c}{Model} \\
\hline
Number & $\mu$Hz & Star & Model & $n$ & $l$ & $m$ \\ \hline
1  & 2641.99 & 378.5026 & (374) &(5)&(3)&(0)         \\
2  & 2837.60 & 352.4105 & 352.2550 & 0 & 0 & 0  \\
3  & 2850.83 & 350.7750 & 350.7777 & 1 & 1 & -1 \\
4  & 2877.16 & 347.5650 & 347.3991 & 1 & 1 & 0  \\
5  & 2906.28 & 344.0825 & 344.0850 & 1 & 1 & +1 \\ \hline
\end{tabular}
\begin{tabular}{|l|c|c|c|c|}\hline
 & Mass & H shell mass & $T_{\rm eff}$ & $\log g$ \\ \hline
Spectroscopy &  & & $29500\pm 300$K & $5.50\pm 0.05$ \\ \hline
Model & 0.4725 & 0.0025 & 29635 & 5.518\\ \hline
\end{tabular}
\end{table}

\subsection{Testing the mode identifications}

A test of our (or any) model is the measured constraint on rotational
velocity. The observed average rotational splitting of 27.7$\mu$Hz imposed on our
$\ell =1$ model identification implies a rotation period of 0.42 days (10
hours). With a radius of 0.20~R$_{\odot}$, this model has an equatorial
rotation velocity of 24~km\, s$^{-1}$. To match the constraints of HRW
($v\sin i\leq 5km\cdot s^{-1}$) 
requires $i\leq 12^{\circ}$. If we use the less restrictive value
determined by HRW for only the unblended spectral lines in Feige~48 of
$v\sin i\leq 10km\cdot s^{-1}$, our inclination limit increases to $i\leq
25^{\circ}$.

The identification of an $\ell=1$ triplet and a radial mode suggest an
observation that can be used to test the model.  As in white dwarfs (Kepler
et al. 2000), time series spectroscopy (particularly in the ultraviolet)
should present an effective means of discerning between low and high order
($\ell$) nonradial pulsations. Though it is still in its infancy for sdBV
stars (O'Toole et al. 2002; Woolf et al. 2002), \emph{if} the 378\,s
period is $\ell$=3, it should be obvious in UV spectroscopy (perhaps less
so in the optical) where it should have a significantly higher amplitude
than at optical wavelengths. The same is true for our identification of the
$\ell$=1 triplet. If any member of our identified triplet really has a
different $\ell$ value, the wavelength dependence of its amplitude will be
different. 
Such a test should be obtained as an 
independent confirmation of our  $\ell =1$ determination.
While this test can be applied to any sdBV star, Feige~48 has
comparatively long pulsation periods (exceeded only by PG~1605+072) and its
rather simple temporal spectrum (only 5 periods compared to 55 for
PG~1605+072) make it an ideal candidate for time series spectroscopic
study.

\section{Stability of the pulsation periods}
As a star evolves, the pulsation properties evolve in response.  In the case 
of the subdwarf B stars, evolutionary 
models indicate that they reside on or near the ZAHB for 
approximately $10^8$ years.  Upon exhausting 
their core helium supply, they
 leave the HB, their $\log g$ goes down, and pulsation
 periods lengthen.  The time 
scale for evolutionary pulsation
 period changes (\.{P}/P) after leaving the HB is about 10 times 
faster than while  on the HB.

If  Feige~48  has a comparatively small log $g$ because 
it is a mature HB star that has left the ZAHB, we  
expect Feige~48 to have an evolutionary
\.{P} smaller than PG~1605+072, yet larger than for 
shorter period pulsators. Since PG~1605+072 does not appear to have pulsations stable
enough for an analysis of secular period change caused by evolution 
(Reed 2001),
Feige~48 is the best candidate to examine the $e$-folding time for structural
changes caused by its core evolution.
As a guide, the model described in Section 4 has a \.{P}=$1.714\cdot
10^{-5}$~s\,yr$^{-1}$.
 With $\sim$3 years of usable data,
the phase of a 350\,s period should change by $\sim$14 seconds in that
time. This is close to our limit of detection.

\begin{figure*}
\centerline{ \epsfig{figure=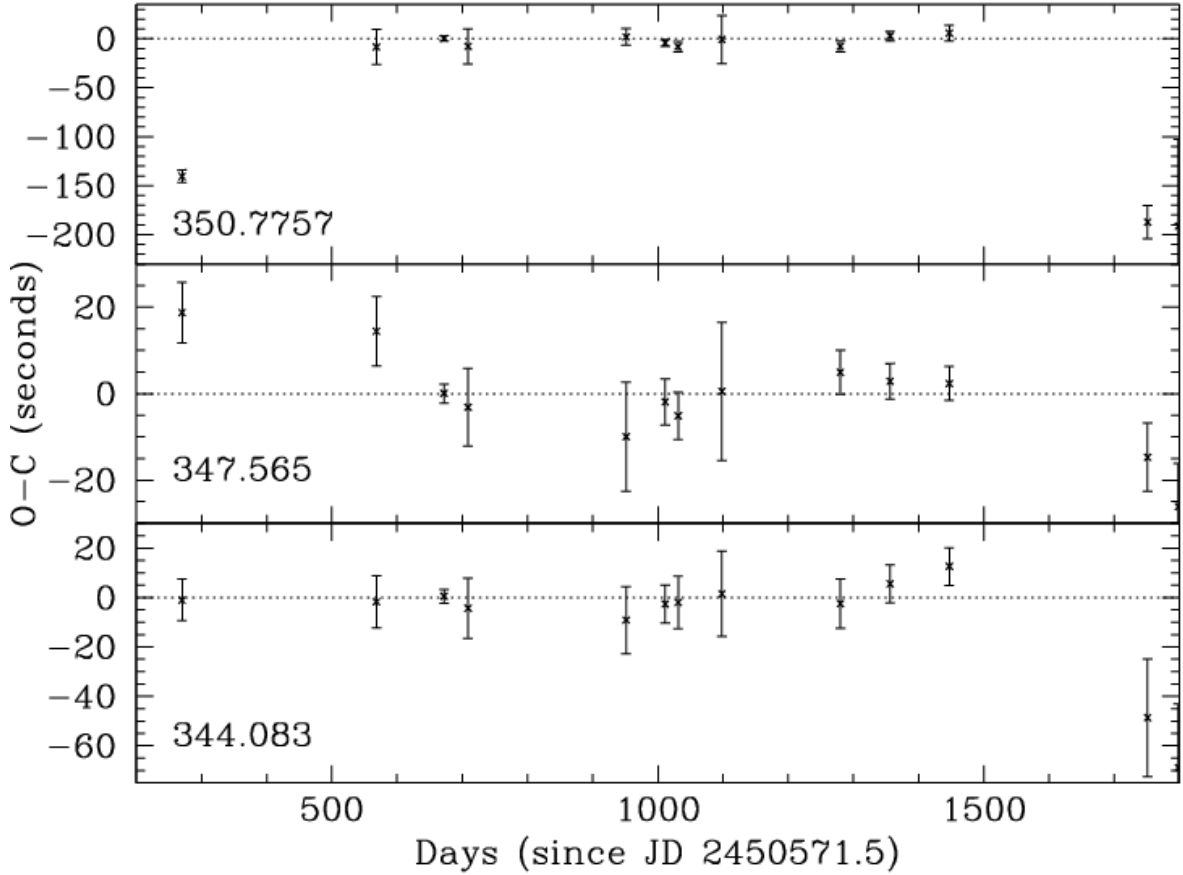,width=6.5in} }
\caption
{O-C diagrams for the 3 largest amplitude modes.
\label{fig04} }
\end{figure*}

To examine long-term phase change, we followed methods outlined in
Winget et al. (1985) and Costa \& Kepler (2000).
First, we obtain a best-fitting least squares fit to all of the periodicities
present over the entire span of the observations (Table ~\ref{tab04}).  
We then fix the
frequencies at these best-fitting values, and recompute the pulsation phases
(again via least squares) for each
 group in Table~\ref{tab02} (note that Groups III and V were
divided into two subgroups each because of the long length of the runs). 
This computed phase represents the observed time of maximum ($O$) for that
group, which differs from the computed time ($C$) from the fit to all data.
The 
resulting $O-C$ diagram
is shown in Fig.~\ref{fig04} for the three highest amplitude modes
(with the pulsation 
period indicated in each panel). The phase zero point is that defined 
in K98 as JD=2450571.50. 

Fig.~\ref{fig04} shows that the phases are stable throughout most
of our observations. Until the Group X data were collected, we believed the
Group I data suffered a timing error (proposed in K98). However,
it now seems apparent the pulsation modes were only stable over a limited
timespan (from Group II through IX). As a check, we analyzed various 
subgroups of Group X data and reproduced the same phase result.
Such a problem is observed in other sdBV stars over a much shorter timescale
(Reed 2001). Though we are disappointed in the apparent lack of phase
stability, we still have an approximately three year span of phase-stable 
observations. We therefore used the phase-stable data to place upper
limits on the magnitude of  \.{P}/P.
The combined $O-C$ data of the three
highest amplitude modes were weighted and fitted using least squares and
give \.{P}/P=$4.9\pm 5.3\times 10^{-16}s^{-1}$.
The data are therefore consistent with zero period change
and provide a $1 \sigma$ lower limit on an evolutionary timescale of
$3.1\times 10^7$ years. Of course evolution is not the only thing that
can drive period changes (see for example Papar\'{o} et al. 1998). However,
evolutionary models predict that sdBV periods should change in a predictable
way (increasing just off the ZAHB, then decreasing after core helium
exhaustion, and finally increasing
again during shell helium fusion). By measuring \.{P} for several sdBV stars
at different stages, we should be able to determine if evolution (as
predicted) is driving the period changes.

\subsection{$O-C$ variations from reflex orbital motion - planets around
Feige~48?}
\begin{figure*}
\centerline{
\epsfig{figure=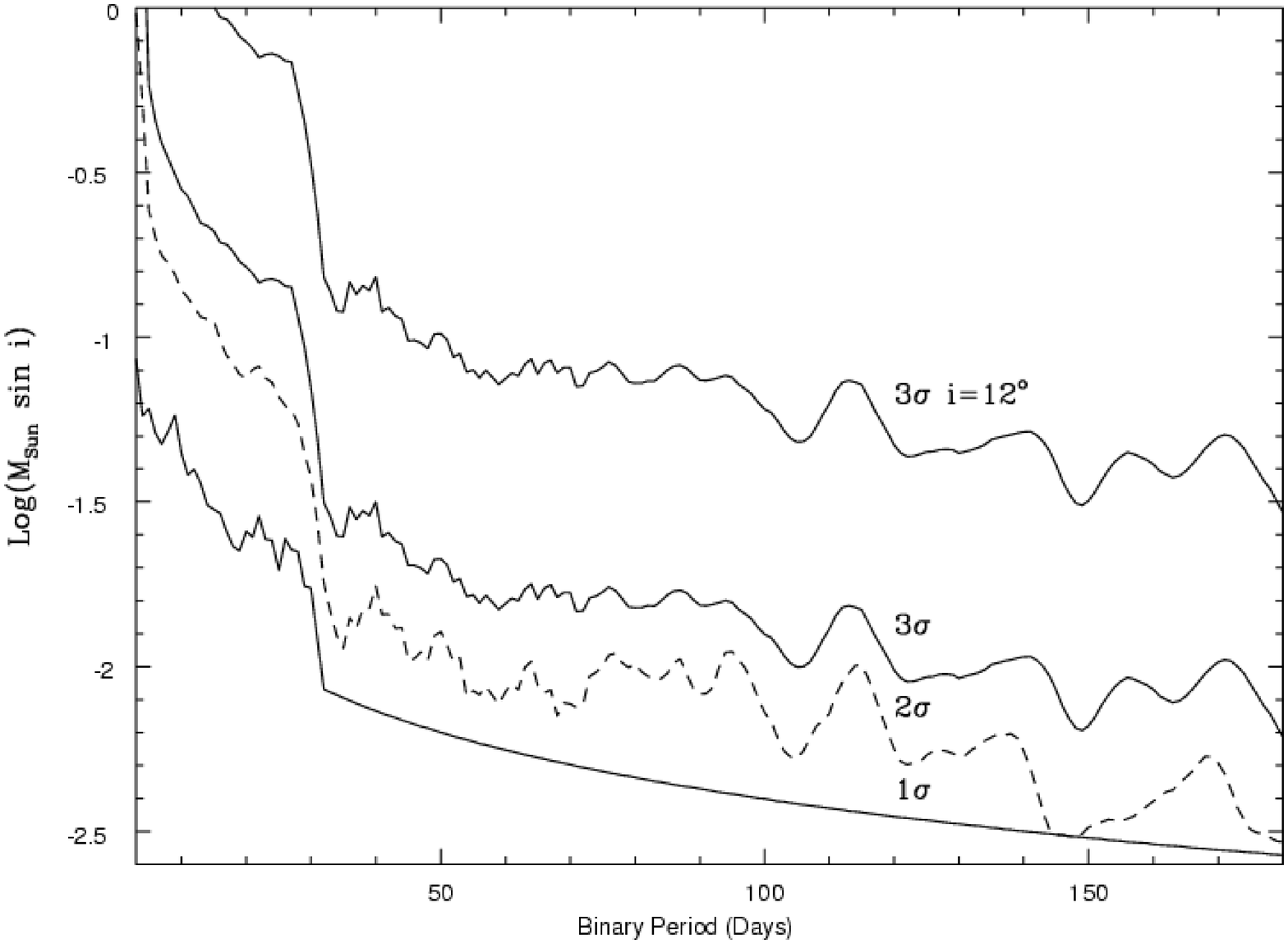,width=3.5in} \epsfig{figure=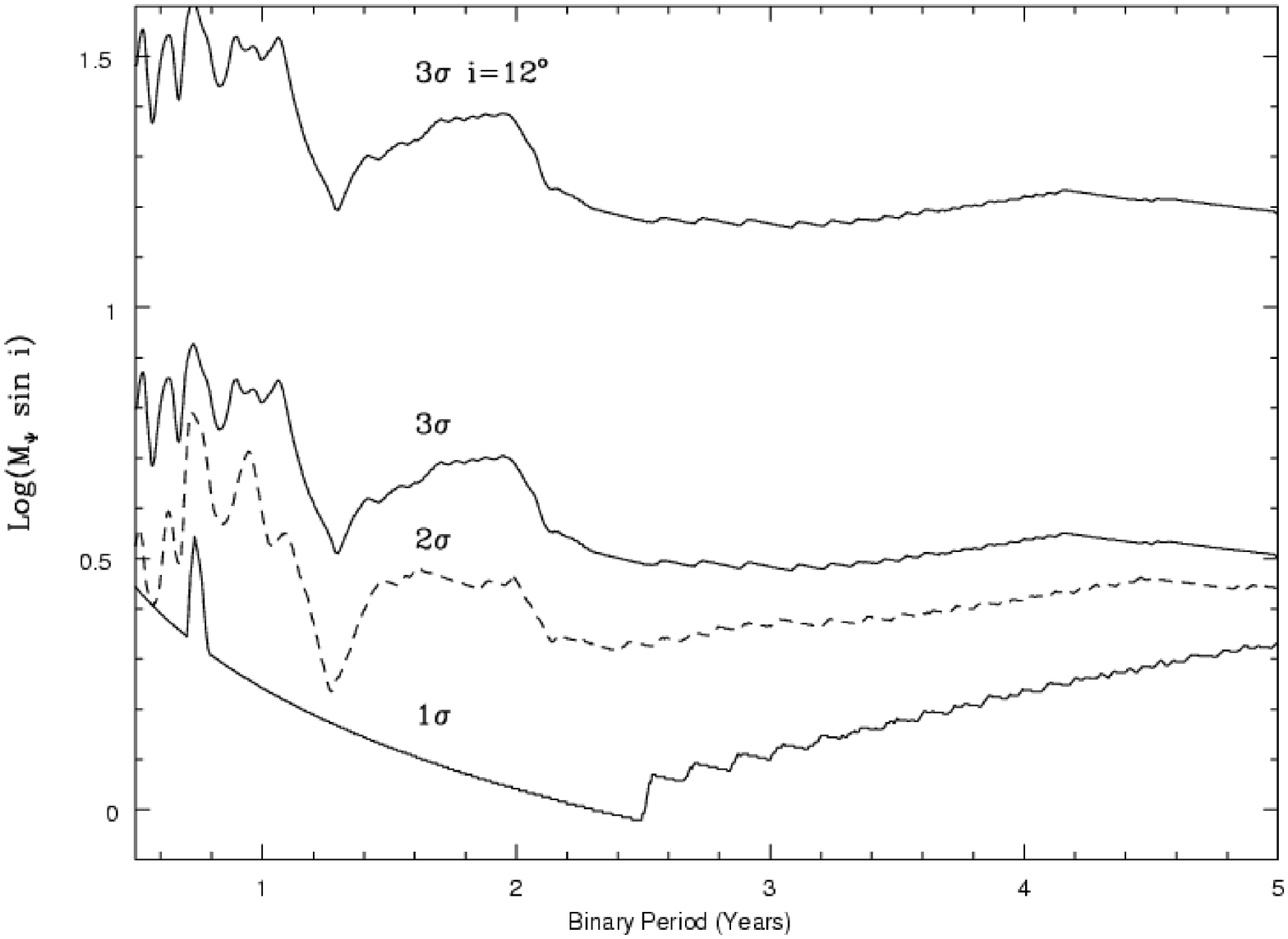, width=3.5in} }
\caption{1$\sigma$, 2$\sigma$ (dashed line), and 3$\sigma$ upper limits on 
companions to Feige~48. The time axis is continuous between panels,
but changes scale from days to years. The mass axis is discontinuous
between panels as the left panel is in solar masses and the right
panel has units of Jupiter masses.}
\label{planet}
\end{figure*}

With the phase-stable portion of the data,
we can also place useful limits on companions to
Feige~48. Orbital reflex motion would create a periodic shift in the
pulsation's arrival time, which would be observed in pulsation
phase. Thus any companion must create a periodic phase change within our $O-C$
uncertainties
over a scale of days to
 years\footnote{We assume we could detect an orbital period up
to twice our observed time base.}. To place limits on companions,
we calculated companion mass as a function of binary period and
semimajor axis by fitting
sine curves (for circular orbits) within our 1, 2, and 3$\sigma$ $O-C$ limits.
To assure an upper limit (M$\sin i$) on reflex motion, 
we use the ``noise'' in the FT of our $O-C$ as a
1$\sigma$ lower limit: This results in a minimum phase shift of 5 seconds
for binary periods under 20 days and 4 seconds for longer periods.
Fig.~\ref{planet} graphically presents our sensitivity to orbital companions.
The top line represents the 3$\sigma$ limit for
$i=12^o$ (our model constraint).
In the short period
case (periods under 30 days), the constraint is the limit
imposed by the phase errors of individual runs within
the data; in the long period case
it is the flatness of the $(O-C)$ diagram (including the errors of combined
runs) over the phase stable region of our observations.
The drop at 30 days corresponds to
the change from
$O-C$ values determined for single runs to group data sets.

Feige~48 is a horizontal branch star that has lost considerable
mass between the red giant branch and its current evolutionary state. Any
companion separated by more than $\sim$1 AU is
far enough away that common--envelope evolution (or vaporisation) has
been avoided.  In addition, the orbital separation will have roughly
doubled as Feige~48 lost approximately half its mass during
the red giant phase.
This should produce two cases for binaries: Close
stellar companions with original separations $\leq 1$AU
will produce a short period  binary
(which have periods on order of weeks or less) after a common
envelope phase. Companions distant enough to avoid a common
envelope phase (or vaporisation) will have orbital periods on
the order of a year or more. The two panels of Fig.~\ref{planet}
reflect this duality (though it does cover all periods between 2
days and 5 years).

The left panel of Fig.~\ref{planet} indicates our limits on stellar
companions. Our 1$\sigma$ limit is less than 0.1M$_{\odot}\sin i$ for a
binary period of 3 days. The right panel shows our limits on
sub--stellar companions. Our ``average'' 3$\sigma$ limit for $i=12^o$
is $\approx 12$M$_{\psi}$, while our best 1$\sigma$ limit
 would detect Jupiter at a
period of 2.5 years.
Our data are currently
 sensitive to the  extra solar ``warm Jupiter'' type planets being
detected\footnote{A complete list of extra solar planets is maintained
at http://www.obspm.fr/encycl/catalog.html.}
at a distance of 0.6 - 3.0 AU. Planets with
orbital separations less than $\sim$1 AU would not
have survived the red giant phase.
Our data do not rule out a companion in an extremely short period binary or at
low inclination.

\section{Conclusions}
From our multi-season photometry of Feige~48, we have consistently
detected five pulsation periods. Of these five, three ($f$1, $f$3 and
$f$4) are consistent with K98. One frequency
($f$5) differs by a daily alias, while K98's fifth
frequency (2874~$\mu$Hz) is not detected in our data. In data
sets V and IX, we also detect new pulsation frequencies at 2890
and 2843~$\mu$Hz respectively,
indicating that there may be some stochastically excited pulsations in 
Feige~48. This is consistent with pulsation behavior seen in another sdBV star,
PG~1605+072 (Reed 2001).

Our attempted model fit follows the
strategy that has been successfully applied to other classes of pulsating
stars, but has rarely worked for sdBV stars; namely using observed
frequency splittings to impose $\ell$ constraints on models.  Using standard
evolutionary models, our preliminary model grid includes a model that is able
to explain all but the lowest frequency stable pulsations. Most
exciting is the fact that such standard models fail to explain all
five frequencies.  Thus, despite its relative simplicity and the richness
of the parameters available, the failure of this model suggests that
standard stellar evolution theory does not fully explain the evolution of
sdB stars and\/ or the nature of pulsations within them. We have something
new to learn.

Our modeling example shows that 
Feige~48 should also serve as an interesting test for other methods
of mode identification. Though optical multicolour photometry was not
useful for identifying pulsation modes in KPD2109+4401(Koen 1998), 
we expect that UV
multicolour photometry as
developed by Robinson, et al. (1995),
will be useful to determine \emph{if} high--order
$\ell$ modes are present in sdBV stars (as indicated by Brassard et al. 2001
and Bill\`eres et al. 2000). The argument for high--order $\ell$ values
is particularly interesting in light of the frequencies detected in Group
V's data. If the lowest frequency mode is disregarded, the remaining modes
have frequency spacings of 13.3, 26.4, 12.8, and 16.3~$\mu$Hz respectively.
If these were all parts of a single, rotationally split mode, it would require
$\ell\geq 3$. Such an $\ell$ value should be apparent
in UV multicolour photometry (see, for example, Kepler et al 2000). As such, we
look forward to the analysis of HST data obtained by Heber (2002).
Should Heber's (2002) HST data agree with our $\ell$=1 interpretation,
Feige~48 would make an excellent star to calibrate other mode identification
methods in sdBV stars such as optical time-series spectroscopy (O'Toole et al
2002; Woolf et al 2002).

 The results of our $O-C$ analysis are
consistent with a non-binary nature for the star within the  data limits.
It also indicates that using the $O-C$ diagram to detect planets around
evolved stars is possible, though in this case we did not detect any.
We plan to continue to monitor Feige~48 over the 
next several years to tighten the constraints on planetary companions.

Our limit on a stellar companion also addresses the origin of sdB stars.
Binary evolution is a candidate for producing sdB stars, either through 
common-envelope evolution 
(Sandquist, Taam \& Burkert 2000; Green,
Liebert \& Saffer 2000) or via Roche-lobe overflow near the tip of
the red giant branch (Green, Liebert \& Saffer 2000). Though
observations indicate that a great many sdB stars \emph{are} in binaries
(Reed \& Steining 2003; Green, Liebert \& Saffer 2000; Han et al 2002), either
the evolutionary sequence that produces sdB stars is independent of binary
evolution (D'Cruz et al 1996), is bimodal, or has several paths
that can result in the production of an sdB star (perhaps including the merger
of two low mass white dwarfs as described by Iben \& Tutukov 1986). For
the case of Feige~48, it would appear that it is either in a short period
binary (whose orbital period is commensurate with the $\sim$10 hour
rotation period predicted with our model), in a long period binary with an 
orbital period substantially longer than our data (which would rule out
Roche-lobe overflow, so the companion would have no effect on the
evolution of the pre-sdB star), in a binary with an extremely low
inclination or a single star.

\section{acknowledgments}

M.D. Reed was partially funded by NSF grant AST9876655 and by the NASA
Astrophysics Theory Program through grant NAG-58352 and would like to
thank the McDonald Observatory TAC for generous time allocation.
S. Dreizler, S.L. Schuh and J.L. Deetjen (Visiting Astronomers at the
German-Spanish Astronomical Centre, Calar Alto, operated by the
Max-Planck-Institut for Astronomy, Heidelberg, jointly with the 
Spanish National Commission for Astronomy) acknowledge travel grant
DR~281/10-1 from the Deutsche Forschungsgemeinschaft. 
P. Moskalik is supported in part by Polish KBN grant 5~P03D~012~020.
A. Ulla acknowledges financial support from the Spanish Ministry of Science
and Tecnology under grant AYA2000-1691. The IAC80 0.8m telescope is
operated in the Spanish Observatorio del Teide (Tenerife) by the Instituto
de Astrof?sica de Canarias.

\end{document}